\documentclass[a4paper,12pt]{article}
\usepackage{amsmath}
\usepackage{accents}
\usepackage{bm}
\usepackage[english]{babel}
\usepackage{graphicx}
\usepackage[usenames]{color}
\usepackage{colortbl}
\usepackage{dsfont}
\usepackage{mathtools}
\usepackage{commath}
\usepackage[font=footnotesize,figurewithin=none]{caption}
\usepackage{qcircuit}
\usepackage[caption=false]{subfig}
\usepackage[toc,page,title]{appendix}

\begin{document}

\centerline {\LARGE{Detecting the purely imaginary Fisher zeros}}
\centerline {\LARGE{of an Ising spin system}}
\centerline {\LARGE{on a quantum computer}}
\medskip
\centerline {A. R. Kuzmak$^1$, V. M. Tkachuk$^2$}
\centerline {\small \it E-Mail: $^1$andrijkuzmak@gmail.com, $^2$voltkachuk@gmail.com}
\medskip
\centerline {\small \it Department for Theoretical Physics, Ivan Franko National University of Lviv,}
\medskip
\centerline {\small \it 12 Drahomanov St., Lviv, UA-79005, Ukraine}

\date{\today}
\begin{abstract}
We propose a protocol for studying the purely imaginary Fisher zeros of the Ising model on a quantum computer. Our protocol is based on the direct relation between the partition function for purely imaginary
temperature and the evolution operator of the Ising model. In this case, the inverse temperature is equal to the time of evolution. This protocol allows one to measure the zeros
only those localized on the imaginary axes. We test this protocol on the ibm-lagos quantum computer in the cases of a 3-spin chain and a triangle cluster in a purely imaginary magnetic field, as well as a 7-spin cluster
in which the interaction between spins reproduces the architecture of the quantum computer.
\end{abstract}

\section{Introduction \label{sec1}}

A novel approach for investigating the thermodynamic properties of the ferromagnetic Ising model in a magnetic field was introduced by Lee and Yang in their papers \cite{LeeYangZeros1, LeeYangZeros}.
This method centers on identifying the partition function zeros, commonly known as Lee-Yang zeros, in the complex plane of the magnetic field. These zeros play a crucial role in defining the analytic properties
of free energy and examining the nature of phase transitions. Lee and Yang established a theorem affirming that these zeros lie on the unit circle in the complex plane \cite{LeeYangZeros}. Subsequently, this theorem was extended
to other systems \cite{genforarbspin1, genforarbspin2, genforarbspin3, genforarbint1, genforarbint2, genforarbint3, genforarbint4, LeeYangZerosahi}. Fisher proposed an analytical study of the partition function
by determining its zeros in relation to the complex inverse temperature called Fisher zeros \cite{FisherZeros}.

The pioneering experimental detection of Lee-Yang zeros is described in \cite{binek1998, binek2001}. The authors established a link between the density function of zeros on the Lee-Yang circle and the
isothermal magnetization of the Ising ferromagnet ${\rm FeCl_2}$ in an axial magnetic field. Subsequent research \cite{zerospartfuncspin3, zerospartfuncspin4} delved into the relationship between
Lee-Yang zeros in the Ising ferromagnet and the decoherence of the probe spin. In this context, the detection of Lee-Yang zeros was transposed into the time domain, which allows one to provide the direct experimental observation
of them on the trimethylphosphite molecule \cite{zerospartfuncspin1}. These investigations provide a direct way for the experimental study of partition function zeros. Inspired by these seminal studies,
numerous recent investigations in this field have emerged. As a result, papers
\cite{kuzmak2019, zerospartfunbose2, zerospartfunbose4, zerospartfunbose5, Brange2023, zerospartfuncspin5, zerospartfuncspin6, zerospartfuncspin7, kuzmak20192, Deger2019, Deger2020, Deger20202, Tao2022} have been published,
exploring the relationships between partition function zeros and the thermodynamic properties, as well as critical parameters, of various many-body systems. As an illustrative example, we generalized the relationship between
Lee-Yang zeros and observables of a probe system for an arbitrarily high spin Ising ferromagnet \cite{kuzmak2019}.

In recent years, the exploration of the behavior of many-body quantum systems has become feasible with the advent of quantum computers. Krishnan et al. first introduced a protocol for detecting partition function zeros
of the Ising model on quantum computers \cite{Krishnan2019}. This protocol relies on implementing simple two-qubit gates, local spin rotations, and projective measurements along two orthogonal quantization axes.
Additionally, ancilla qubits are required, the number of which depends on the graph underlying the Ising model. A more comprehensive protocol for detecting partition function zeros of arbitrary models on quantum computers
was given by Francis et al. \cite{Francis2021}. Remarkably, this protocol necessitates only one ancilla qubit to calculate partition function zeros on a universal quantum computer. The detection of partition
function zeros occurs through the measurement of this qubit. The authors conducted experiments, measuring the zeros of the partition function of the XXZ spin chain on a trapped-ion quantum computer and quantum circuit simulators,
as the model transitions from Ising-like model to XY-like model. In a recent publication \cite{Laba2023}, the study of the partition function of the Ising model on a graph, using a quantum computer, was presented. The authors proposed simulating
the Boltzmann factor as a trace of some evolution operator with an effective Hamiltonian over ancilla spins (qubits) corresponding to graph links.

Our approach is based on paper \cite{Heyl2013}, which explores the connection between the Fisher zeros of the partition function
and the overlap amplitude of the unitary evolution of a given initial quantum state with itself. We propose a protocol for studying the purely imaginary Fisher zeros of the Ising model on a quantum computer.
Unlike previous proposed protocols utilizing ancilla qubits, our method relies on the direct relationship between
the partition function for purely imaginary temperature and the evolution operator of the Ising model (Section~\ref{protocol}). In this context, the inverse temperature is equivalent to the time of evolution.
It is important to note that this protocol is limited to measuring zeros exclusively on the imaginary axes, precluding measurements in the complex plane. To validate our approach, we apply out protocol on the IBM quantum computer,
in the case of a 3-spin chain (Subsection~\ref{isingsysthreespins}), triangle cluster (Subsection~\ref{isingsysthreespinscirk}) subjected to a purely imaginary magnetic field, and a 7-spin cluster where
the interaction between spins reproduces the architecture of the quantum computer (Subsection~\ref{isingsysseven}). Conclusions are presented in Sec.~\ref{conc}.

\section{Protocol for detecting purely imaginary Fisher zeros on a quantum computer \label{protocol}}

We consider a system of $N$ spin-$1/2$ particles described by the Ising model in a magnetic field:
\begin{eqnarray}
H=\frac{1}{4}\sum_{i,j}^NJ_{ij}\sigma_i^z\sigma_j^z+\frac{h}{2}\sum_i^N\sigma_i^z,
\label{isingham}
\end{eqnarray}
where $J_{ij}$ represents the interaction couplings between spins, $h$ is proportional to the magnitude of the magnetic field,
and $\sigma_i^z$ is the $z$-th Pauli matrix for the $i$-th spin. Using the results from paper \cite{Laba2023}, we calculate the partition function of the Ising spin system without the need for ancilla qubits.
The approach in that earlier work involves implementing the Boltzmann factor and computing the partition function of the Ising model on a quantum computer. Since the inverse temperature is a real quantity in that case,
the Boltzmann factor is a non-unitary operator. Consequently, the authors propose a method that requires an ancilla qubit to implement it. In contrast, our aim is to investigate the Fisher zeros of the partition function
at a purely imaginary temperature. In this case, the Boltzmann factor becomes a unitary operator, which allows it to be implemented directly on a quantum computer without the need for ancilla qubits.
The thermodynamic equilibrium state of this system is described by the Boltzmann distribution, and the corresponding density operator is given by:
\begin{eqnarray}
\rho=e^{-\beta H}/Z(\beta)=e^{-\beta \left(\frac{1}{4}\sum_{i,j}^NJ_{ij}\sigma_i^z\sigma_j^z+\frac{h}{2}\sum_i^N\sigma_i^z\right)}/Z(\beta),
\label{isingdensityop}
\end{eqnarray}
where $Z(\beta)={\rm Tr}\left(e^{-\beta H}\right)$ is the partition function of the system, $\beta=1/T$ is the inverse temperature. Here, we use the units where the Planck and Boltzmann constants are: $\hbar=1$, $k_B=1$.
The partition function of the system can be expressed as follows
\begin{eqnarray}
&&Z(\beta)=2^N\langle {\pmb +}\vert e^{-\beta H}\vert {\pmb +}\rangle\nonumber\\
&&=2^N\langle {\pmb +}\vert e^{-\beta \left(\frac{1}{4}\sum_{i,j}^NJ_{ij}\sigma_i^z\sigma_j^z+\frac{h}{2}\sum_i^N\sigma_i^z\right)}\vert {\pmb +}\rangle,
\label{isingpartfunc}
\end{eqnarray}
where $\vert{\pmb +}\rangle=\vert ++\ldots +\rangle$ represents the state of the system defining the positive direction of all spins along the $x$-axis. The state of a single spin is expressed
as $\vert+\rangle=1/\sqrt{2}\left(\vert 0\rangle + \vert 1\rangle\right)$, and $\vert 0\rangle$, $\vert 1\rangle$ are the eigenstates of the $\sigma^z$ Pauli operator corresponding to positive and negative eigenvalues, respectively.

We are interested in scenarios where the inverse temperature is a complex with only an imaginary part. We investigate the values of this imaginary temperature for which the partition function becomes zero.
This observation allows us to simulate the identification of purely imaginary Fisher zeros by evolving the spin system on a quantum computer. We achieve this by transforming $\beta = i\alpha$, where $\alpha$ is a real number.
Subsequently, the density operator (\ref{isingdensityop}) takes the form of the evolution operator with $\alpha$ serving as the temporal parameter \cite{Heyl2013}. Taking into account this fact and the relation $\vert +\rangle = \textsf{H}\vert 0 \rangle$,
the partition function (\ref{isingpartfunc}) is expressed as:
\begin{eqnarray}
&&Z(\alpha)=2^N\langle {\bf 0}\vert \textsf{H}^{\otimes N} e^{-i\alpha \left(\frac{1}{4}\sum_{i,j}J_{ij}\sigma_i^z\sigma_j^z+\frac{h}{2}\sum_i\sigma_i^z\right)} \textsf{H}^{\otimes N}\vert {\bf 0}\rangle\nonumber\\
&&=2^N\langle {\bf 0}\vert e^{-i\alpha \left(\frac{1}{4}\sum_{i,j}J_{ij}\sigma_i^x\sigma_j^x+\frac{h}{2}\sum_i\sigma_i^x\right)}\vert {\bf 0}\rangle,
\label{isingpartfunc2}
\end{eqnarray}
where $\vert{\bf 0}\rangle=\vert 00\ldots 0\rangle$ represents the state of the system, defining the positive direction of all spins along the $z$-axis, and $\textsf{H}$ is the Hadamard operator.
For the second part of equation (\ref{isingpartfunc2}), we use the relation $\sigma^x=\textsf{H}\sigma^z\textsf{H}$ and the fact that all terms in the Hamiltonian (\ref{isingham}) mutually commute. The determination of zeros in the partition function
is thus reduced to finding values of $\alpha$ at which mean (\ref{isingpartfunc2}) becomes zero.

\begin{figure}[!!h]
\centerline{\includegraphics[scale=0.7, angle=0.0, clip]{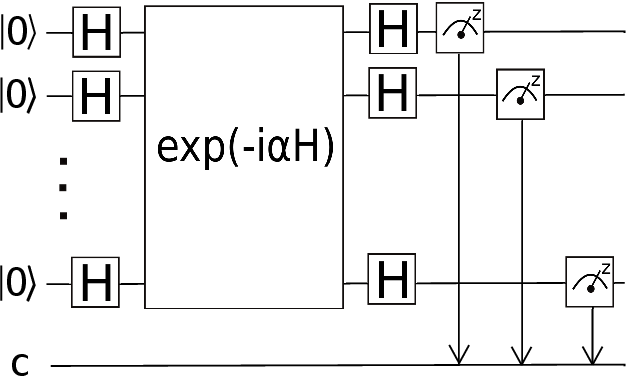}}
\caption{The protocol designed to measure the Fisher zeros of the spin-1/2 Ising model defined by Hamiltonian (\ref{isingham}). Here $\textsf{H}$ is the Hadamard operator.}
\label{protocolgr}
\end{figure}

It is noteworthy that the mean value of the evolution operator with the XX-like Ising model is calculated over the state $\vert{\bf 0}\rangle$. This simplifies the process of finding Fisher zeros on a quantum computer as only
measurements on the $\vert{\bf 0}\rangle$ state need to be considered. The quantum computer, after measurement, provides the square of the modulus of $Z(\alpha)$ (\ref{isingpartfunc2}). The values of $\alpha$
at which the partition function and its modulus become zero coincide. Therefore, by measuring the square of the modulus of $Z(\alpha)$ on a quantum computer, we can determine the values of $\alpha$ at which it equals zero.
These values correspond to purely imaginary Fisher zeros. The protocol outlining these calculations is depicted in Fig.~\ref{protocolgr}.

Since all terms in the Hamiltonian (\ref{isingham}) mutually commute, the evolution operator $\exp(-i\alpha H)$ is expressed as a product of separate operators $e^{-i\alpha \frac{J_{ij}}{4}\sigma_i^z\sigma_j^z}$ and $e^{-i\alpha \frac{h}{2} \sigma_i^z}$.
Quantum circuits representing these operators are shown as follows:
\begin{eqnarray}
&&e^{-i\frac{J_{ij}}{4}\alpha \sigma_i^z\sigma_j^z}= \Qcircuit @C=1em @R=.7em {
&\lstick{i}&\ctrl{1} & \qw & \ctrl{1} & \qw \\
&\lstick{j}& \targ &\gate{{\sf R_z(J_{ij}\alpha/2)}} & \targ & \qw
}\label{isingqc}\\
&&e^{-i\alpha \frac{h}{2}\sigma_i^z}=
\Qcircuit @C=1em @R=.7em {
&\lstick{i} &\gate{{\sf R_z(h\alpha)}}&\qw
},
\label{isingqc2}
\end{eqnarray}
where $R_z(\phi)$ represents the rotation of the qubit state around the $z$ axis by the angle $\phi$, and:
\begin{eqnarray}
\Qcircuit @C=1em @R=.7em {&\lstick{i}&\ctrl{1} & \qw \\ &\lstick{j}& \targ & \qw}\nonumber
\end{eqnarray}
depicts the controlled-NOT gate, performing a ${\rm NOT}$ on the $j$th qubit whenever the $i$th qubit is in state $\vert 1\rangle$. The fact that the quantum circuit in equation (\ref{isingqc}) satisfies the corresponding
evolution operator can be easily proved. On the one hand, the action of the evolution operator on the basis states of two spins is as follows
\begin{eqnarray}
&&e^{-i\frac{J_{ij}}{4}\alpha \sigma_i^z\sigma_j^z}\vert 00\rangle_{ij} =e^{-i\frac{J_{ij}}{4}\alpha}\vert 00\rangle_{ij},\quad e^{-i\frac{J_{ij}}{4}\alpha \sigma_i^z\sigma_j^z}\vert 01\rangle_{ij} =e^{i\frac{J_{ij}}{4}\alpha}\vert 01\rangle_{ij},\nonumber\\
&&e^{-i\frac{J_{ij}}{4}\alpha \sigma_i^z\sigma_j^z}\vert 10\rangle_{ij} =e^{i\frac{J_{ij}}{4}\alpha}\vert 10\rangle_{ij},\quad e^{-i\frac{J_{ij}}{4}\alpha \sigma_i^z\sigma_j^z}\vert 11\rangle_{ij} =e^{-i\frac{J_{ij}}{4}\alpha}\vert 11\rangle_{ij}.\nonumber
\end{eqnarray}
On the other hand, for example, let us sequentially apply the operators that are in the quantum circuit to the state $\vert 10\rangle_{ij}$
\begin{eqnarray}
&&\textsf{CNOT}_{ij}{\textsf{R}_z}_j(J_{ij}\alpha/2))\textsf{CNOT}_{ij}\vert 10\rangle_{ij}=\textsf{CNOT}_{ij}{\textsf{R}_z}_j(J_{ij}\alpha/2))\vert 11\rangle_{ij}\nonumber\\
&&=e^{i\frac{J_{ij}}{4}\alpha}\textsf{CNOT}_{ij}\vert 11\rangle_{ij}=e^{i\frac{J_{ij}}{4}\alpha}\vert 10\rangle_{ij}.\nonumber
\end{eqnarray}
We obtain the result as in the case of the direct action of the evolution operator on this state. In the same way, it is possible to verify the validity of this quantum circuit for other basis states.
We now study the Fisher zeros of specific spin models on a quantum computer.

\section{Measurement of purely imaginary Fisher zeros on the ibm-lagos quantum computer \label{isingint}}

\begin{figure}[!!h]
\centerline{\includegraphics[scale=0.4, angle=0.0, clip]{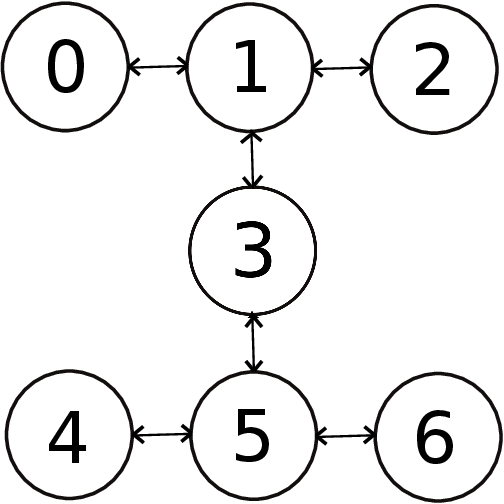}}
\caption{The structure of the ibm-lagos quantum computer.}
\label{ibmq-lima}
\end{figure}

In this section, we apply our protocol to calculate the purely imaginary Fisher zeros of certain Ising models on the ibm-lagos quantum computer. The ibm-lagos quantum device is composed of seven superconducting qubits
that interact with each other, as illustrated in Fig.~\ref{ibmq-lima}. Access to this device is freely available through the IBM Q Experience cloud service \cite{IBMQExp}. Quantum circuits can be executed on these computers
using various quantum gates, including the controlled-NOT gate, $\sigma^x$ Pauli operator, $\sqrt{\sigma^x}$ gate, and the $R_z(\phi)$ gate, corresponding to rotating the qubit around the $z$-axis
by the angle $\phi$ \cite{OpenQasm}. The interaction between qubits is facilitated through controlled-NOT gates, as indicated by the bidirectional arrow in Fig.~\ref{ibmq-lima}.
Each pair of qubits connected by the bidirectional arrow can be driven by the controlled-NOT operator in a way that each of the qubits can be both a control and a target.

We investigate purely imaginary Fisher zeros for the 3-spin and 7-spin Ising models with various interaction configurations between spins. Utilizing the ibm-lagos quantum computer, we explore the impact of the magnetic
field and interaction configuration on the presence of purely imaginary Fisher zeros in these systems. Note that in paper \cite{Laba2023}, the partition function was calculated for a two-spin Ising model without a magnetic field.
In contrast, we extend the analysis by including a magnetic field and investigate its impact on the appearance of Fisher zeros in the partition function. Let us consider into the details of these cases.

\subsection{The 3-spin chain \label{isingsysthreespins}}

\begin{figure}[!!h]
\centerline{\subfloat[]{\label{threespinschain}}\includegraphics[scale=0.75, angle=0.0, clip]{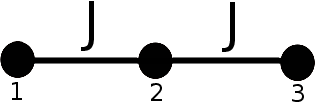}\qquad
\subfloat[]{\label{threespinstriangle}}\includegraphics[scale=0.75, angle=0.0, clip]{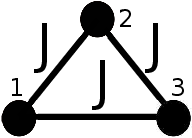}}
\caption{The structure of the 3-spin chain and triangle cluster. The interaction between spins is defined by value $J$.}
\label{threespins}
\end{figure}

In this subsection, we investigate the Fisher zeros of a three-spin chain, as illustrated in Fig.~\ref{threespinschain}. The Hamiltonian for this model is given by:
\begin{eqnarray}
H=\frac{J}{4}\left(\sigma_1^z\sigma_2^z+\sigma_2^z\sigma_3^z\right)+\frac{h}{2}\left(\sigma_1^z+\sigma_2^z+\sigma_3^z\right).
\label{threespinsham}
\end{eqnarray}
Here, we assume that the interaction between spins is uniform and equals $J$. The eigenstates and corresponding eigenvalues of this Hamiltonian are:
\begin{align}
&\vert\psi_1\rangle = \vert 000\rangle , && E_{1}=\frac{J}{2}+\frac{3}{2}h,\nonumber\\
&\vert\psi_2\rangle = \vert 001\rangle , && E_{2}=\frac{h}{2},\nonumber\\
&\vert\psi_3\rangle = \vert 010\rangle , && E_{3}=-\frac{J}{2}+\frac{h}{2},\nonumber\\
&\vert\psi_4\rangle = \vert 100\rangle , && E_{4}=\frac{h}{2},\nonumber\\
&\vert\psi_5\rangle = \vert 011\rangle , && E_{5}=-\frac{h}{2},\nonumber\\
&\vert\psi_6\rangle = \vert 101\rangle , && E_{6}=-\frac{J}{2}-\frac{h}{2},\nonumber\\
&\vert\psi_7\rangle = \vert 110\rangle , && E_{7}=-\frac{h}{2},\nonumber\\
&\vert\psi_8\rangle = \vert 111\rangle , && E_{8}=\frac{J}{2}-\frac{3}{2}h.
\label{eigenvaleigenstatechain}
\end{align}
Now, let us calculate the partition function of the model defined by Hamiltonian (\ref{threespinsham}).
Substituting eigenstates and eigenvalues (\ref{eigenvaleigenstatechain}) into the first line of equation (\ref{isingpartfunc2}), we obtain the partition function of a 3-spin chain
\begin{eqnarray}
&&Z(\alpha)=\sum_{i=1}^8\langle \psi_i \vert \exp{\left(-i\alpha E_i\right)} \vert \psi_i \rangle\nonumber\\
&&=2\left(e^{-iJ\alpha/2}\cos\left(3h\alpha/2\right)+e^{iJ\alpha/2}\cos\left(h\alpha/2\right)\right.\nonumber\\
&&\left.+2\cos\left(h\alpha/2\right)\right).
\label{isingpartfunc3spins}
\end{eqnarray}
\begin{figure}[!!h]
\centerline{\includegraphics[scale=0.5, angle=0.0, clip]{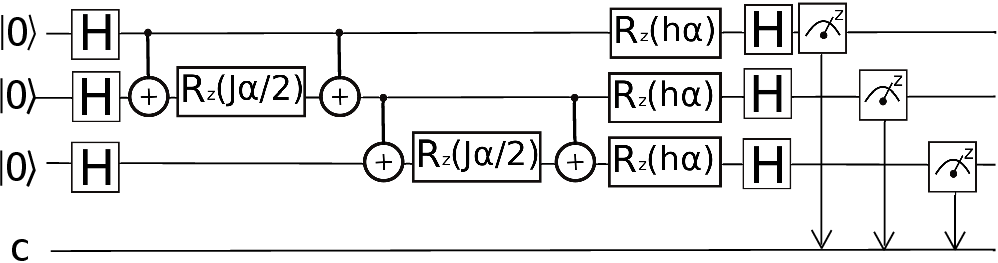}}
\caption{The protocol which allows one to measure the Fisher zeros of the three spin-1/2 system defined by Hamiltonian (\ref{threespinsham}).}
\label{protocol_tree_spins}
\end{figure}
\begin{figure}[!!h]
\subfloat[]{\label{threespinspartfuncabsencemf}}\includegraphics[scale=0.515, angle=0.0, clip]{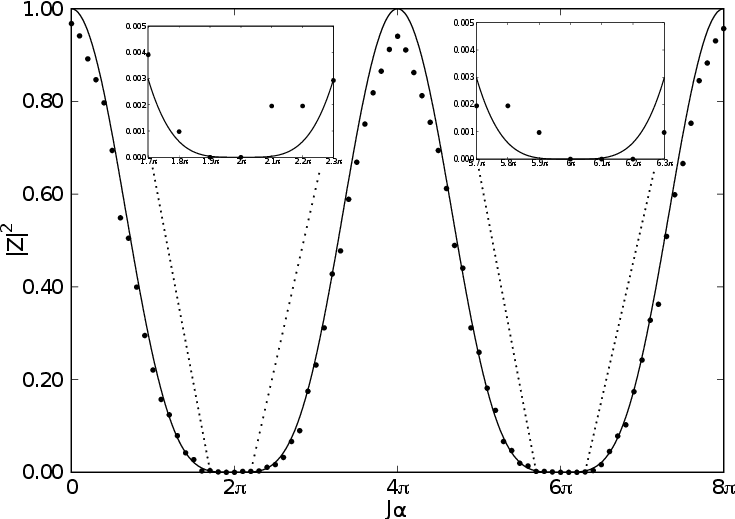}\qquad
\subfloat[]{\label{threespinspartfuncmfle}}\includegraphics[scale=0.515, angle=0.0, clip]{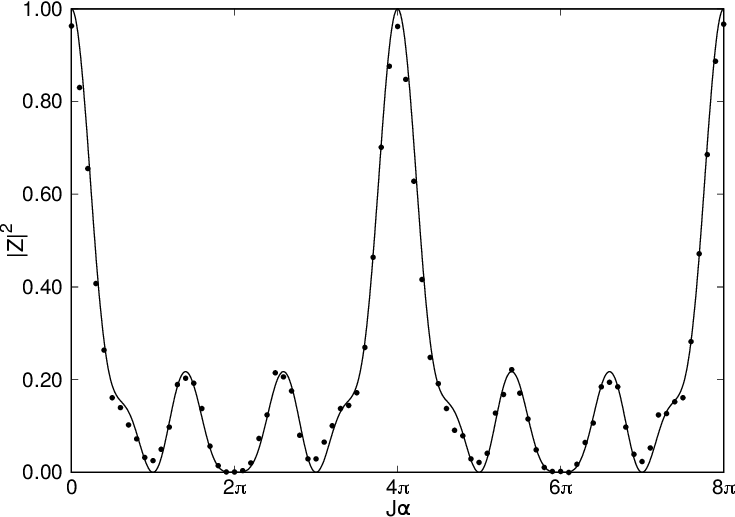}
\caption{The plot depicts the square of the modulus of the partition function (\ref{isingpartfunc3spins}) for the 3-spin chain as a function of the $\alpha$.
The two cases presented include $h=0$ (a) and $h=J$ (b), where $J$ denotes the coupling constant. The solid lines represent the theoretical predictions, while the dots correspond to the results obtained on the ibm-lagos quantum computer.
Fisher zeros are identified at specific points, such as $J\alpha=2\pi$ and $6\pi$ for the case with $h=0$ (a), and $J\alpha=\pi$, $2\pi$, $3\pi$, $5\pi$, $6\pi$, $7\pi$ for the case with $h=J$ (b). Here we
divided the partition function by 8.}
\label{threespinspartfunc}
\end{figure}
In this expression, we observe purely imaginary Fisher zeros induced by the interaction between spins and the magnetic field. In the case of $h=0$, zeros appear at the points $J\alpha = 2(2n+1)\pi$, where $n\in\mathds{Z}$.
In the presence of a magnetic field, regardless of the interaction, zeros appear concerning the field at the points $h\alpha=(2n+1)\pi$. The symmetrical terms in the system's spectrum (\ref{eigenvaleigenstatechain}) contribute
to the emergence of purely imaginary Fisher zeros in the partition function (\ref{isingpartfunc3spins}). These terms differ by sign, leading to certain temperature values where terms in the partition function become zero.
As indicated in equations (\ref{eigenvaleigenstatechain}) and (\ref{isingpartfunc3spins}), even in the absence of a magnetic field ($h=0$), purely imaginary Fisher zeros are present. This is due to the symmetrical nature
of the spectrum in this case.

From partition function (\ref{isingpartfunc3spins}), we obtain the equation which determines the zeros in the $J-h$ plane. These zeros are ones for imaginary $\alpha$ and real magnetic field.
For this purpose, the square of the modulus of partition function (\ref{isingpartfunc3spins}) should be put to zero $\vert Z(\alpha)\vert^2=0$. Then the equation which describes the connection between
the values of interaction and magnetic field in the case of Fisher zeros reads
\begin{eqnarray}
&&\cos\left(\frac{J\alpha}{2}\right)=\left[-\cos\left(\frac{h\alpha}{2}\right)\left(\cos\left(\frac{h\alpha}{2}\right)+\cos\left(\frac{3h\alpha}{2}\right)\right)\right.\nonumber\\
&&\left.\pm \sqrt{\cos\left(\frac{h\alpha}{2}\right)\left(\cos\left(\frac{h\alpha}{2}\right)-\cos\left(\frac{3h\alpha}{2}\right)\right)^3}\right]\nonumber\\
&&\text{\LARGE $/$}\left[2\cos\left(\frac{h\alpha}{2}\right)\cos\left(\frac{3h\alpha}{2}\right)\right].
\label{leeyang3spinschain}
\end{eqnarray}

Using the method outlined in the previous section, we employ the ibm-lagos quantum computer to detect the purely imaginary Fisher zeros of the three-spin system defined by Hamiltonian (\ref{threespinsham}).
We examine these zeros in two cases: when there is no magnetic field ($h=0$) and when $J=h$. In the first scenario, the partition function depends on the parameter $J\alpha$ as $Z(\alpha)=8\cos^2\left(J\alpha/4\right)$.
In the second case, we use the partition function (\ref{isingpartfunc3spins}) with $h=J$. By varying $J\alpha$ in the range from $0$ to $8\pi$ with a step of $\pi/10$, we measure $\vert Z(\alpha)\vert^2$ on the quantum computer.
For each value of $J\alpha$, we perform 1024 measurements of $\vert Z(\alpha)\vert^2$ on the quantum computer.

In Fig.~\ref{protocol_tree_spins}, the protocol for detecting the purely imaginary Fisher zeros of a three-spin chain in a magnetic field, defined by Hamiltonian (\ref{threespinsham}), is presented.
To replicate the behavior of $\vert Z(\alpha)\vert^2$, the results measured on the state $\vert 000\rangle$ should be taken into account.

In Fig.~\ref{threespinspartfunc} we present results for the cases with $h=0$ and $h=J$. As observed, in the case of $h=0$, the measured results align very well with the theoretical ones. However, in the case of $h=J$,
the quantum gates defining the magnetic field's influence on each spin (\ref{isingqc2}) introduce additional errors in the protocol, slightly compromising the results obtained on the quantum computer. Nevertheless,
the figures indicate that the measured purely imaginary Fisher zeros are accurately defined within the range covering $0.1\pi-0.2\pi$.

\subsection{The 3-spin triangle cluster \label{isingsysthreespinscirk}}

Here, we investigate the purely imaginary Fisher zeros of the 3-spin triangle cluster illustrated in Fig.~\ref{threespinstriangle}.
In contrast to the previous case, the Hamiltonian here includes an additional term describing the interaction between the first and third spins:
\begin{eqnarray}
H=\frac{J}{4}\left(\sigma_1^z\sigma_2^z+\sigma_2^z\sigma_3^z+\sigma_1^z\sigma_3^z\right)+\frac{h}{2}\left(\sigma_1^z+\sigma_2^z+\sigma_3^z\right).
\label{threespinshamtriangl}
\end{eqnarray}
The eigenstates and corresponding eigenvalues of this Hamiltonian are given by:
\begin{align}
&\vert\psi_1\rangle = \vert 000\rangle , && E_{1}=\frac{3}{4}J+\frac{3}{2}h,\nonumber\\
&\vert\psi_2\rangle = \vert 001\rangle , && \nonumber\\
&\vert\psi_3\rangle = \vert 010\rangle , && E_{2}=-\frac{J}{4}+\frac{h}{2},\nonumber\\
&\vert\psi_4\rangle = \vert 100\rangle , && \nonumber\\
&\vert\psi_5\rangle = \vert 011\rangle , && \nonumber\\
&\vert\psi_6\rangle = \vert 101\rangle , && E_{3}=-\frac{J}{4}-\frac{h}{2},\nonumber\\
&\vert\psi_7\rangle = \vert 110\rangle , && \nonumber\\
&\vert\psi_8\rangle = \vert 111\rangle , && E_{4}=\frac{3}{4}J-\frac{3}{2}h.
\label{eigenvaleigenstatetriangl}
\end{align}
Similar to the previous case, we calculate the partition function of the triangle spin cluster defined by Hamiltonian (\ref{threespinshamtriangl}). This partition function is expressed as:
\begin{eqnarray}
Z(\alpha)=2\left(e^{-i3J\alpha/4}\cos\left(3h\alpha/2\right)+3e^{iJ\alpha/4}\cos\left(h\alpha/2\right)\right).\nonumber\\
\label{isingpartfunc3spinstriangle}
\end{eqnarray}
In the absence of a magnetic field ($h=0$), no purely imaginary Fisher zeros are observed. However, the presence of a magnetic field leads to their appearance. This is due to symmetric terms with respect to $h$
in the spectrum (\ref{eigenvaleigenstatetriangl}), which are transformed into cosines in equation (\ref{isingpartfunc3spinstriangle}). Consequently, points $h\alpha=(2n+1)\pi$ ($n\in\mathds{Z}$) correspond to Fisher zeros.
In this case the equation which defines the zeros of partition function (\ref{isingpartfunc3spinstriangle}) in the $J-h$ plane takes the form
\begin{eqnarray}
\cos(J\alpha)=-\frac{\cos^2\left(\frac{3h\alpha}{2}\right)+9\cos^2\left(\frac{h\alpha}{2}\right)}{6\cos\left(\frac{3h\alpha}{2}\right)\cos\left(\frac{h\alpha}{2}\right)}.
\label{leeyang3spinsclaster}
\end{eqnarray}

\begin{figure}[!!h]
\centerline{\includegraphics[scale=0.55, angle=0.0, clip]{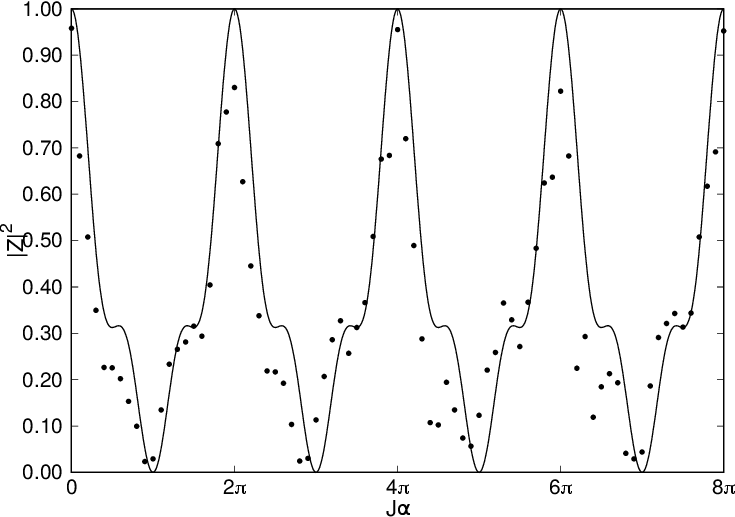}}
\caption{The square of the modulus of the partition function (\ref{isingpartfunc3spinstriangle}) is plotted as a function of $J\alpha$ in the case of $h=J$. The solid lines represent the theoretical predictions,
while the dots depict the results obtained on the ibm-lagos quantum computer. The purely imaginary Fisher zeros correspond to the points at $J\alpha=\pi$, $3\pi$, $5\pi$, and $7\pi$. Here we
divided the partition function by 8.}
\label{threespinscirkpartfunc}
\end{figure}

We compute $\vert Z(\alpha)\vert^2$ as a function of $J\alpha$ on the ibm-lagos quantum computer in the case of $h=J$. In addition to the previous case, here the quantum circuit has
one more quantum gate (\ref{isingqc}) that describes the Ising interaction between the first and third spins. Although a quantum computer (see Fig.~\ref{ibmq-lima}) does not have any set of three qubits that interact in a triangle,
however, the interaction between the two outermost qubits is generated through the middle qubit. In this case, the compiler converts the given program according to the capabilities of the computer.
It uses more quantum gates than it would in the case of a triangular connection between qubits. In Fig.~\ref{threespinscirkpartfunc}, we illustrate the dependence of $\vert Z(\alpha)\vert^2$
as a function of $J\alpha$ within the range from $0$ to $8\pi$. On the quantum computer, measurements are performed with a step of $\pi/10$. For each value $J\alpha$, we perform 1024 measurements of $\vert Z(\alpha)\vert^2$.
The obtained results are then compared with theoretical predictions.
It is evident that at the specified points $J\alpha=\pi$, $3\pi$, $5\pi$, and $7\pi$, the measured values of $\vert Z(\alpha)\vert^2$ exhibit minima that are close to zero. However, due to gate and readout errors, they do not reach zero.
The additional quantum gates that define the interaction between the first and third spins induce errors that more deviate the obtained results away from the theoretical prediction.

\subsection{The 7-spin cluster \label{isingsysseven}}

Finally, we determine the purely imaginary Fisher zeros of the Ising model, where the spins reproduce the structure of the ibm-lagos quantum computer. In other words, the interaction between spins is determined by the connections between
qubits, as shown in Fig.~\ref{ibmq-lima}. We also assume that the magnetic field is absent. If we denote the spins by the numbers of qubits on the quantum computer, then the Hamiltonian takes the form
\begin{eqnarray}
H=\frac{J}{4}\left(\sigma_0^z\sigma_1^z+\sigma_1^z\sigma_2^z+\sigma_1^z\sigma_3^z+\sigma_3^z\sigma_5^z+\sigma_4^z\sigma_5^z+\sigma_5^z\sigma_6^z\right).\nonumber\\
\label{sevenspinsham}
\end{eqnarray}
The partition function of this system has the form
\begin{eqnarray}
&&Z(\alpha)= 2^7\cos^6\left(\frac{J\alpha}{4}\right).
\label{isingpartfunc7spins}
\end{eqnarray}
The Fisher zeros correspond to values $J\alpha=2(2n+1)\pi$, where $n\in\mathds{Z}$.

\begin{figure}[!!h]
	\centerline{\includegraphics[scale=0.55, angle=0.0, clip]{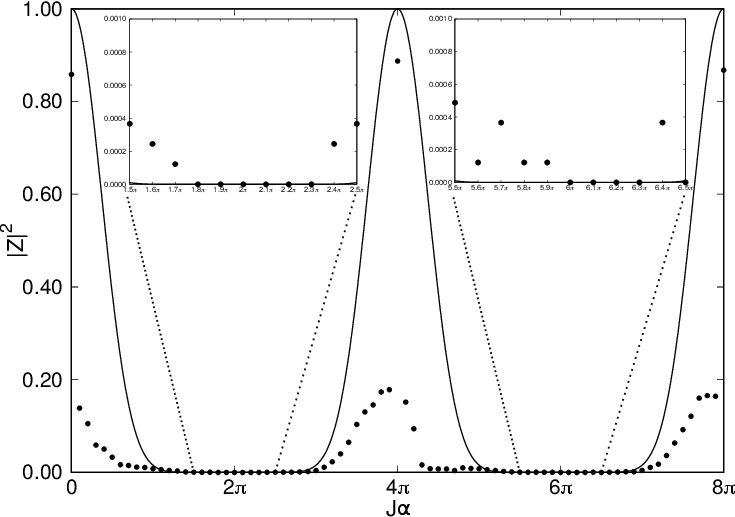}}
	\caption{The square of the modulus of the partition function (\ref{isingpartfunc7spins}) as a function of $J\alpha$. The solid lines represent the theoretical predictions, and the dots depict the results obtained
on the ibm-lagos quantum computer. Purely imaginary Fisher zeros correspond to the points at $J\alpha=2\pi$ and $6\pi$. Here we
divided the partition function by $2^7$.}
	\label{sevenspinspartfunc}
\end{figure}

In the same manner as in the previous subsections, we investigate $\vert Z(\alpha)\vert^2$ on the ibm-lagos quantum device. However, to enhance the accuracy of the measurements, each value of $J\alpha$ is measured 8192 times.
This is due to the fact that, unlike the previous three-spin systems, here we have a 7-spin system. Such a system requires more quantum computer resources, including quantum gates and the number of qubits that are measured.
Consequently, this leads to an increase in errors that accumulate during calculations on a quantum computer. In Fig.~\ref{sevenspinspartfunc}, we present results for $\vert Z(\alpha)\vert^2$ as a function of $J\alpha$.
Due to the larger size of the system, gate and readout errors are more significant. Therefore, the measured results deviate more from the theoretical predictions. As observed in the figures, the measured Fisher zeros are accurately
defined within a range that covers $0.5\pi$.

\section{Conclusions \label{conc}}

We have introduced a protocol for measuring the purely imaginary Fisher zeros of an Ising spin system in a magnetic field using a quantum computer. In this context, in our case, the inverse temperature
consists solely of the imaginary part. The partition function exhibits characteristics of an evolution operator \cite{Heyl2013}, making it amenable to simulation on a quantum computer. We have demonstrated that the partition function
of the ZZ-Ising model can be expressed as an average over a state where all spins align along the positive direction of the $x$-axis. By rewriting the partition function as the mean value of the XX-Ising model
over a state where all spins align along the positive direction of the $z$-axis (\ref{isingpartfunc2}), we streamline the quantum computer calculation, focusing only on the results obtained from this particular state.
Consequently, there exist specific values of the inverse temperature and magnetic field for which the partition function becomes zero. These values can be pinpointed on a quantum computer
by identifying the points where the mean value (\ref{isingpartfunc2}) reaches zero.

Partition function of the Ising model can be presented in the form of some polynomia. According to the fundamental theorem of algebra the information about the zeros allows us to reconstruct the partition function
of the system (see, \cite{LeeYangZeros,FisherZeros,Francis2021,Singh2024}). Thus, having a partition function of the system, we can investigate thermodynamic properties and phase transitions
of the system (for example, see
\cite{LeeYangZeros1,LeeYangZeros,FisherZeros,Heyl2013,Wu2008}). It is clear that there are no phase transitions in a few-body systems, because they appear in the thermodynamic limit.
However, experimentally, the phase transition begin to appear with an increase in the number of particles in the system. Phase transitions cannot be observed for the systems which contein finite number of spins considered in section~\ref{isingint}.
These examples show that the Fisher zeros can be detected in principle on a quantum computer. We have applied our protocol to simulate and measure the zeros
of a 3-spin chain, a triangle cluster in a magnetic field, and a 7-spin cluster (with interactions between spins replicating the architecture
of the ibm-lagos quantum computer). The results obtained were then compared with theoretical predictions. Overall, the measured results closely align with theoretical expectations, accurately identifying the points corresponding
to partition function zeros on the quantum computer.

\section{Acknowledgements}

This work was supported by Project 77/02.2020 from National Research Foundation of Ukraine.

{}

{\bf Author contributions statements}: A.K. and V.T. developed the protocol, performed the calculations on a quantum computer, analyzed the results, and wrote the main text of manuscript.

{\bf Competing Interests statement}: The authors declare that they have no known competing financial interests or personal relationships that could have appeared to influence the work reported in this paper.

{\bf Data availability}: All data generated or analysed during this study are included in this published article.

\end{document}